\documentclass{elsart}
\usepackage{epsfig}
\usepackage{amsmath}
\usepackage{graphics}
\usepackage{color}
\newcommand{\aff}[2]{Dipartimento di Fisica dell'Universit\`a #1 e Sezione INFN, #2, Italy.}
\newcommand{\affd}[1]{Dipartimento di Fisica dell'Universit\`a e Sezione INFN, #1, Italy.}
\def\ifm#1{\relax\ifmmode#1\else$#1$\fi}
\let\cl=\centerline   \def\x{\ifm{\times}}  \def\f{\ifm{\phi}}
\def\DAF{DA$\Phi$NE}     \def\eps{\ifm{\epsilon}}
\def\gam{\ifm{\gamma}} \def\to{\ifm{\rightarrow}}
\def\ab{\ifm{\sim}}  \def\x{\ifm{\times}}
\def\up#1{\ifm{^{#1}}}  
\def\ff{$\phi$--factory}
\def\to{\ifm{\rightarrow}}
  
\def\pt#1,#2,{\ifm{#1\x10^{#2}}}

\makeatletter
\newskip\z@skip \z@skip=0pt plus0pt minus0pt
\def\m@th{\mathsurround=\z@}
\def\ialign{\everycr{}\tabskip\z@skip\halign} 
\def\eqalign#1{\null\,\vcenter{\openup\jot\m@th
  \ialign{\strut\hfil$\displaystyle{##}$&$\displaystyle{{}##}$\hfil
      \crcr#1\crcr}}\,}
\makeatother

\newcommand{\be}{\begin{equation}}
\newcommand{\ee}{\end{equation}}
\newcommand{\bea}{\begin{eqnarray}}
\newcommand{\eea}{\end{eqnarray}}
\newcommand{\bc}{\begin{center}}
\newcommand{\ec}{\end{center}}
\newcommand{\bt}{\begin{tabular}}
\newcommand{\et}{\end{tabular}}
\newcommand{\bfig}{\begin{figure}}
\newcommand{\efig}{\end{figure}}
\newcommand{\bi}{\begin{itemize}}
\newcommand{\ei}{\end{itemize}}
\newcommand{\bleft}{\begin{flushleft}}
\newcommand{\eleft}{\end{flushleft}}
\newcommand{\bright}{\begin{flushright}}
\newcommand{\eright}{\end{flushright}}
\newcommand{\bpage}{\begin{minipage}}
\newcommand{\epage}{\end{minipage}}

\newcommand{\ket}[1]{\ensuremath{\left|{#1}\right>}}

\newcommand{\vect}[1]{\ensuremath{\overrightarrow{#1}}}

\newcommand{\pip}{\ensuremath{\pi^+\,}}
\newcommand{\pim}{\ensuremath{\pi^-\,}}
\newcommand{\piz}{\ensuremath{\pi^0\,}}
\newcommand{\Ks}{\ensuremath{K_S \,}}
\newcommand{\Kl}{\ensuremath{K_L\,}}

\newcommand{\Eta}{\ensuremath{\eta\,}}
\newcommand{\etap}{\ensuremath{\eta'\,}}
\newcommand{\phot}{\ensuremath{\gamma\,}}

\newcommand{\fiv}{\ensuremath{\varphi_V\,}}
\newcommand{\fip}{\ensuremath{\varphi_P\,}}
\newcommand{\dafne}{ DA\ensuremath{\Phi}NE }
\renewcommand{\to}{\ensuremath{\rightarrow}}
\newcommand{\cm}{\ensuremath{\,{\rm cm}}}

\newcommand{\ps}{\ensuremath{\,{\rm ps}}}

\newcommand{\pbinv}{\ensuremath{\,{\rm pb}^{-1}}}
\newcommand{\GeV}{\ensuremath{\,{\rm GeV}}}
\newcommand{\MeV}{\ensuremath{\,{\rm MeV}}}
\newcommand{\fikskl}{\ensuremath{\phi\rightarrow\Ks\Kl}}

\newcommand{\fietag}{\ensuremath{\phi\rightarrow\eta\gamma\;}}
\newcommand{\fietapg}{\ensuremath{\phi\rightarrow\eta'\gamma\;}}
\newcommand{\etapippimpiz}{\ensuremath{\eta\rightarrow\pip\pim\piz}}

\newcommand{\etagg}{\ensuremath{\eta\rightarrow\gamma\gamma\;}}
\newcommand{\etappippimeta}{\ensuremath{\etap\rightarrow\pip\pim\eta}}

\begin{document}

\begin{frontmatter}
\title{
Measurement of $
\Gamma(\phi\rightarrow\eta'\gamma)/\Gamma(\phi\rightarrow\eta\gamma)$ and
the pseudoscalar mixing angle}
\collab{The KLOE Collaboration}
\author[Na] {A.~Aloisio},
\author[Na]{F.~Ambrosino\corauthref{cor}},
\corauth[cor]{Corresponding author.}
\ead{Fabio.Ambrosino@na.infn.it}
\author[Frascati]{A.~Antonelli},
\author[Frascati]{M.~Antonelli},
\author[Roma3]{C.~Bacci},
\author[Frascati]{G.~Bencivenni},
\author[Frascati]{S.~Bertolucci},
\author[Roma1]{C.~Bini},
\author[Frascati]{C.~Bloise},
\author[Roma1]{V.~Bocci},
\author[Frascati]{F.~Bossi},
\author[Roma3]{P.~Branchini},
% ...
\address[Roma2]{\aff{``Tor Vergata''}{Roma}}
\address[Na]{Dipartimento di Scienze Fisiche dell'Universit\`a ``Federico II'' e Sezione INFN,
Napoli, Italy}
\address[Moscow]{Permanent address: Institute for Theoretical and Experimental Physics, Moscow, Russia.}
\address[Frascati]{Laboratori Nazionali di Frascati dell'INFN, Frascati, Italy.}
\address[Roma3]{\aff{``Roma Tre''}{Roma}}
\address[Trieste]{\affd{Trieste}}
\address[Roma1]{\aff{``La Sapienza''}{Roma}}
\address[Lecce]{\affd{Lecce}}
\address[Pisa]{\affd{Pisa}}
\address[Bari]{\affd{Bari}}
\address[Beijing]{Permanent address: Institute of High Energy Physics, CAS, Beijing, China.}
\address[StonyBrook]{Physics Department, State University of New York at Stony Brook, USA.}
\address[Karlsruhe]{Institut f\"ur Experimentelle Kernphysik, Universit\"at Karlsruhe, Germany.}

\address[Energ]{Dipartimento di Energetica dell'Universit\`a ``La Sapienza'', Roma, Italy.}
\address[Virginia]{Physics Department, University of Virginia, USA.}
\begin{abstract}
We have measured the radiative decays $\fietag$, $\fietapg$ selecting
\pip\pim$\phot \phot \phot$ final state in a sample of \ab5\x10\up7
\f-mesons produced at the Frascati \ff\ DA$\Phi$NE. We obtain
$\Gamma(\fietapg)/\Gamma(\fietag)$ =
\pt(4.70\pm0.47\pm0.31),-3,. From this result we derive new accurate values
for the branching ratio BR($\fietapg$) =  \pt(6.10\pm0.61\pm0.43),-5, and 
the mixing angle of pseudoscalar mesons in the flavour basis $\varphi_P=(41.8^{+1.9}_{-1.6})^\circ$.
\end{abstract}
\end{frontmatter}

Radiative decays of light vector mesons to pseudoscalars have been a
source of precious information  since the early days of the quark model
\cite{BeMor65}. They have been studied in the context of chiral
lagrangians by several authors \cite{BraGrauPan}.
The branching ratio (BR) of the decay \fietapg\ is particularly interesting 
since its value can probe the $s\bar{s}$ and gluonium contents
of the \etap\ \cite{Close92} or the amount of nonet symmetry breaking \cite{Ben99}. In particular, the ratio $R$=BR(\fietapg)/BR(\fietag) can be related to the \Eta-\etap\ mixing parameters~\cite{Ros83,BallFreTyt96,BraEsSca99,BraEsSca01,Feld00} and determines the pseudoscalar mixing angle. 
Even for the case of two mixing angles which appears in extended chiral perturbation theory~\cite{KaisLeut98}, as well as from phenomenological
analyses~\cite{EsFre99}, it has been argued that the two mixing
parameters in the flavour basis are equal apart from terms which violate the
Okubo-Zweig-Iizuka (OZI) rule \cite{FeldKroll98,Defazio00}. It is thus possible to
parameterize mixing in a nearly process independent way by just one
mixing angle, \fip. 
The large BR($B\to K\eta'$) value observed~\cite{expBKetap}, as opposed to theoretical predictions~\cite{theoBKetap}, raises also interest~\cite{Kou00} about the gluonium contents of the $\etap$. This can also be tested from a precise determination of BR$(\phi\to\eta'\gamma)$.
The BR(\fietapg) measurements available to date still have rather large
uncertainties~\cite{PDG,SND99,CMD200b}. 
The study of \fietapg\ decays presented in the following, is  based on an
integrated  luminosity of \ab16\pbinv\ corresponding to some $5\times10^7$ $\phi$  decays collected by the KLOE detector~\cite{kloe} at \DAF~\cite{dafne}, the Frascati $e^+e^-$ collider, during the year 2000. All data were taken at a total energy $w=M_{\phi}$.

The KLOE detector consists of a large cylindrical drift chamber (DC), surrounded by a lead-scintillating fibers electromagnetic calorimeter (EmC). A superconducting coil surrounds the EmC and provides a 0.52 T field along the beam axis. The DC~\cite{DriftChamber}, 4 m diameter and 3.3 m long, has 12,582 all-stereo tungsten sense wires and 37,746 aluminum field wires. The chamber shell is made of carbon
fiber-epoxy composite and the gas used is a 90\% helium, 10\% isobutane
mixture. These choices maximize transparency to photons and reduce $K_L\to 
K_S$ regeneration as well as multiple scattering. Momentum resolution is $\sigma(p_{\perp})/p_{\perp}\!\ab\!0.4\%$. Position resolution is $\sigma_{xy}\!\ab\! 150\mu$m and $\sigma_z\!\ab\!$2 mm. Vertices are reconstructed with an accuracy of \ab3 mm.
The EmC~\cite{Calorimeter} is divided into a barrel and two end-caps, for a
total of 88 modules, covers 98\% of the solid angle. The modules are
read out at both ends by photomultipliers. Readout granularity is $\sim
4.4\times 4.4 \cm^2$, for a total of 2,440 ``cells''. Arrival times and positions in three dimensions of energy deposits are determined from the signals at the two ends. Cells close in time and space are grouped into a calorimeter cluster. The cluster energy $E_{CL}$ is the sum of the cell energies. Cluster time $t_{CL}$ and position
$\vec r_{CL}$, are energy weighed averages. Resolutions are $\sigma_E/E= 5.7\%/\sqrt{E(\GeV)}$ and $\sigma_t=57\ps/\sqrt{E(\GeV)}\oplus 50$ ps. The detector trigger~\cite{Trigger} uses calorimeter and chamber information.

To determine $R$ we search for events~\cite{nota}:
\begin{enumerate}
\item $\fietapg$; with $\etap\to\pip\pim\eta$ and $\eta\to\gamma\gamma$
\item $\fietag$; with $\eta\rightarrow\pip\pim\piz$ and $\piz\to\gamma\gamma$
\end{enumerate}
The final state is \pip\pim\gam\gam\gam\ for both reactions. Most
systematics uncertainties therefore approximately cancel in measuring $R$. $\fietag$ decays are easily selected with small 
background and provide a clean control sample for the analysis.
Process 2, about 100 times more abundant than 1, is the main source of background for \fietapg\ events.
Further background is due to:
\begin{enumerate}\setcounter{enumi}{2}
\item \fikskl events with one charged vertex where
at least one photon is not detected and the \Kl is decaying near the interaction
region (IR)
\item $\phi\rightarrow \pip \pim \piz$ events with an additional
photon detected due to accidental clusters or splitting of clusters in the EmC.\\ 
\end{enumerate}
After some trivial cuts to remove radiative Bhabhas and machine background events, we select events satisfying the following cuts:
\begin{itemize}
\item[a)] Exactly three photons with 
$21^{\circ}<\theta_{\gamma}<159^{\circ}$ and $E_{\gamma}>10$ MeV

\item[b)] Opening angle of each photon pair $>18^{\circ}$

\item[c)] A vertex inside the cylindrical region $\sqrt{x^2+y^2}<4\;\cm$; $|z|<8 \;\cm$ with
two opposite charge tracks.
\end{itemize}

A prompt photon is a calorimeter cluster with no track pointing to it and 
$|(t_{CL}-|\vect{r}_{CL}|/c)|< 5\sigma_t(E_{CL})$.
Small angles are excluded to reduce machine background. 
The opening angle cut ensures that fragments of clusters are not counted as separate photons.
At this level the ratio of the two efficiencies is 
$\epsilon_{\etap\gamma}/\epsilon_{\eta\gamma} = 0.9$. 
It is close to one, given the similarities among the two processes.
The residual difference is due to the efficiency in tracking to the origin the pions, because of their different momentum spectra for the two processes.
After this first level selection we perform a kinematic fit requiring
energy-momentum conservation and that photons travel at the speed of light. Particle 
masses are not constrained. We require $\rm{prob}(\chi^2)>1\%$ for for both
processes (1) and (2).

The only additional cuts applied to select
process (2) are a very loose cut on the energy of the radiative photon
(after kinematic fit) and a cut on pion energy endpoint. The radiative photon can be easily
identified being the hardest in the event for process (2).
We require:
\bi
\item[d)] $320 \MeV < E_{\gamma}^{{\rm rad.}} < 400 \MeV$ 
\item[e)] $E_{\pip}+E_{\pim}< 550 \MeV$
\ei

\begin{figure}[h]
\bc
\epsfig{file=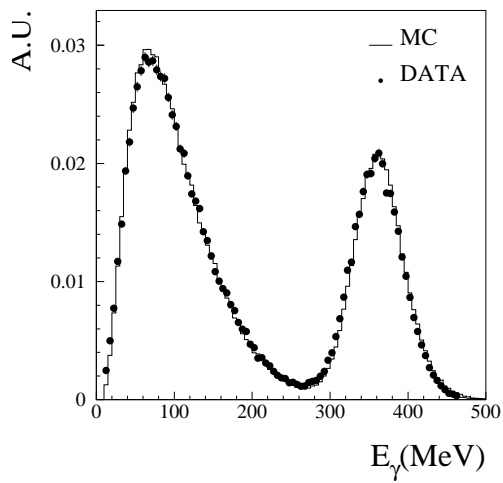,width=7cm}
\caption{{\footnotesize Data - MC comparison for the energy spectrum of
photons in events selected as $\phi \to \eta\gamma \to
\pip\pim\gamma\gamma\gamma$.}}
\label{figspectrum}
\ec
\end{figure}

The first cut is very effective in
reducing residual background from  process (3) where the endpoint for
photon energies is at 280 \MeV. 
The second cut eliminates residual background from process (4). Both cuts
have full efficiency for the signal.
We are then left with $N_{\eta\gamma}$=50210$\pm$220 events. The overall efficiency
for detecting \fietag\ events is evaluated from Monte Carlo simulations (MC) to be 36.5\%. Background is expected to be below 0.5\%, and all observed distributions are in agreement with this estimate.
The abundant and pure  $\fietag$ events are used as a control sample to
evaluate systematic effects on the efficiencies by comparing data and
MC distributions for the variable to which cuts are applied. The
distributions exhibit a remarkable agreement, as shown in
fig.~\ref{figspectrum} for the photons energy spectrum.

\begin{figure}[h]

\cl{\epsfig{file=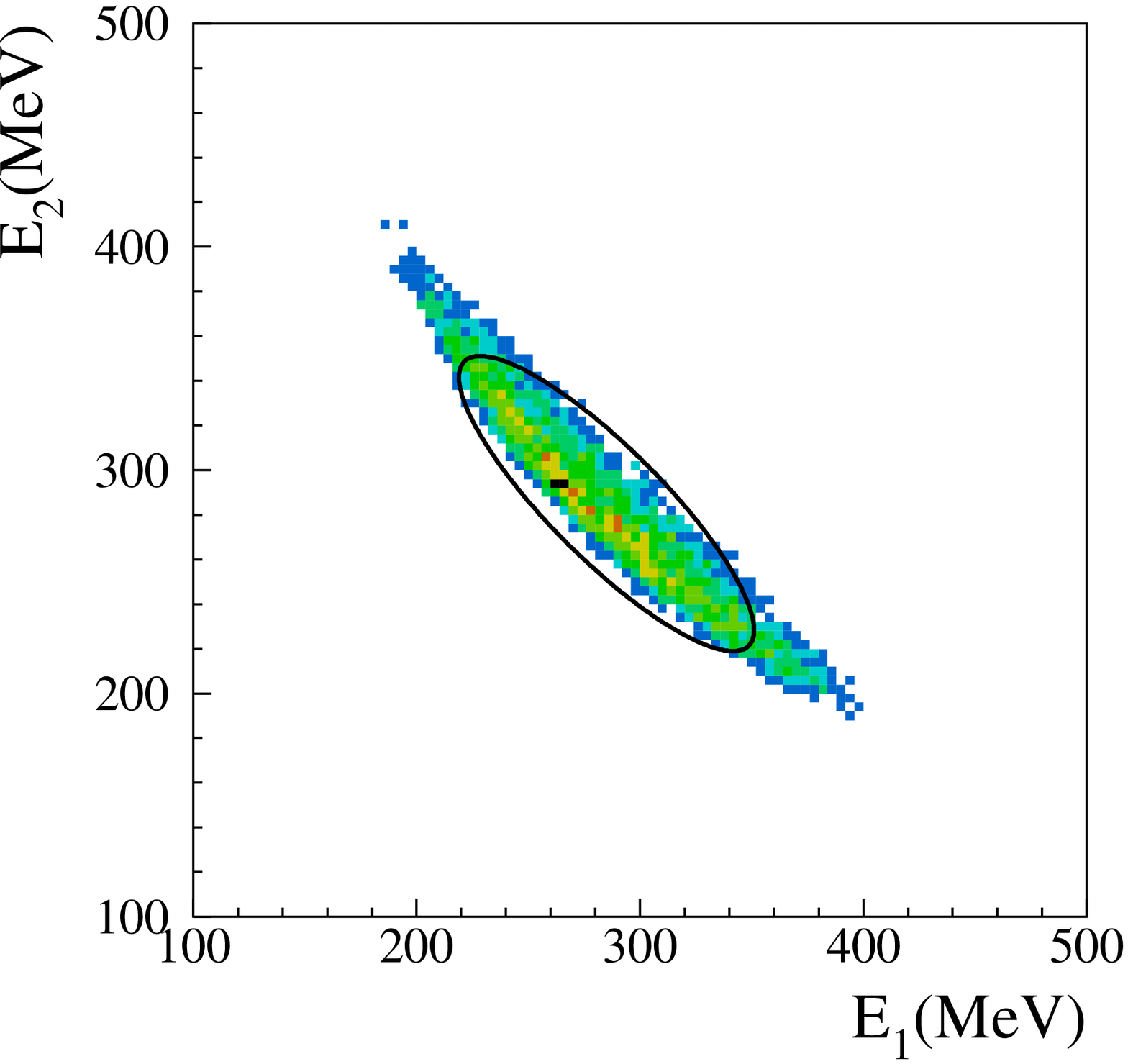,width=7cm}\quad\epsfig{file=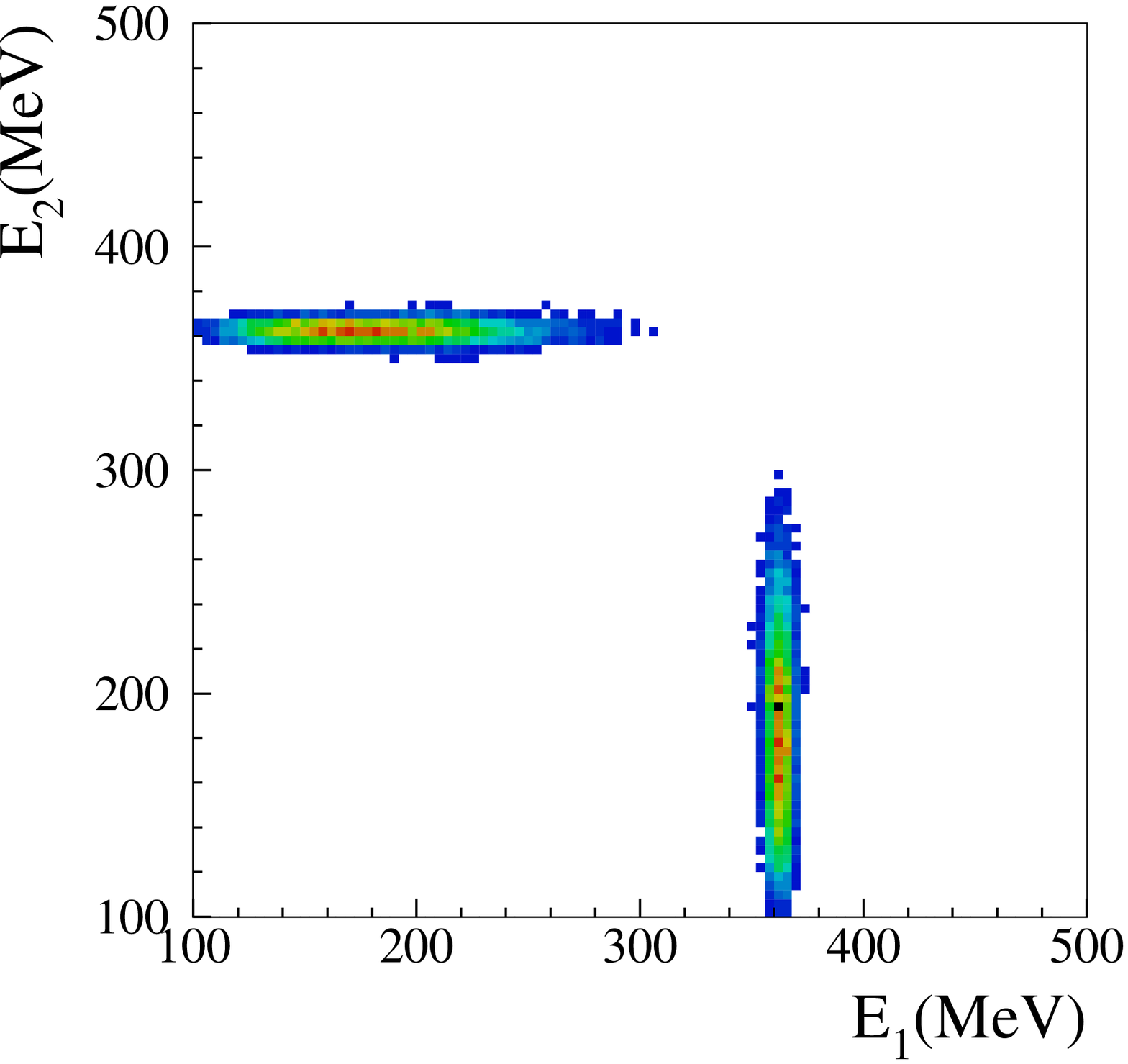,width=7cm}}
\cl{\epsfig{file=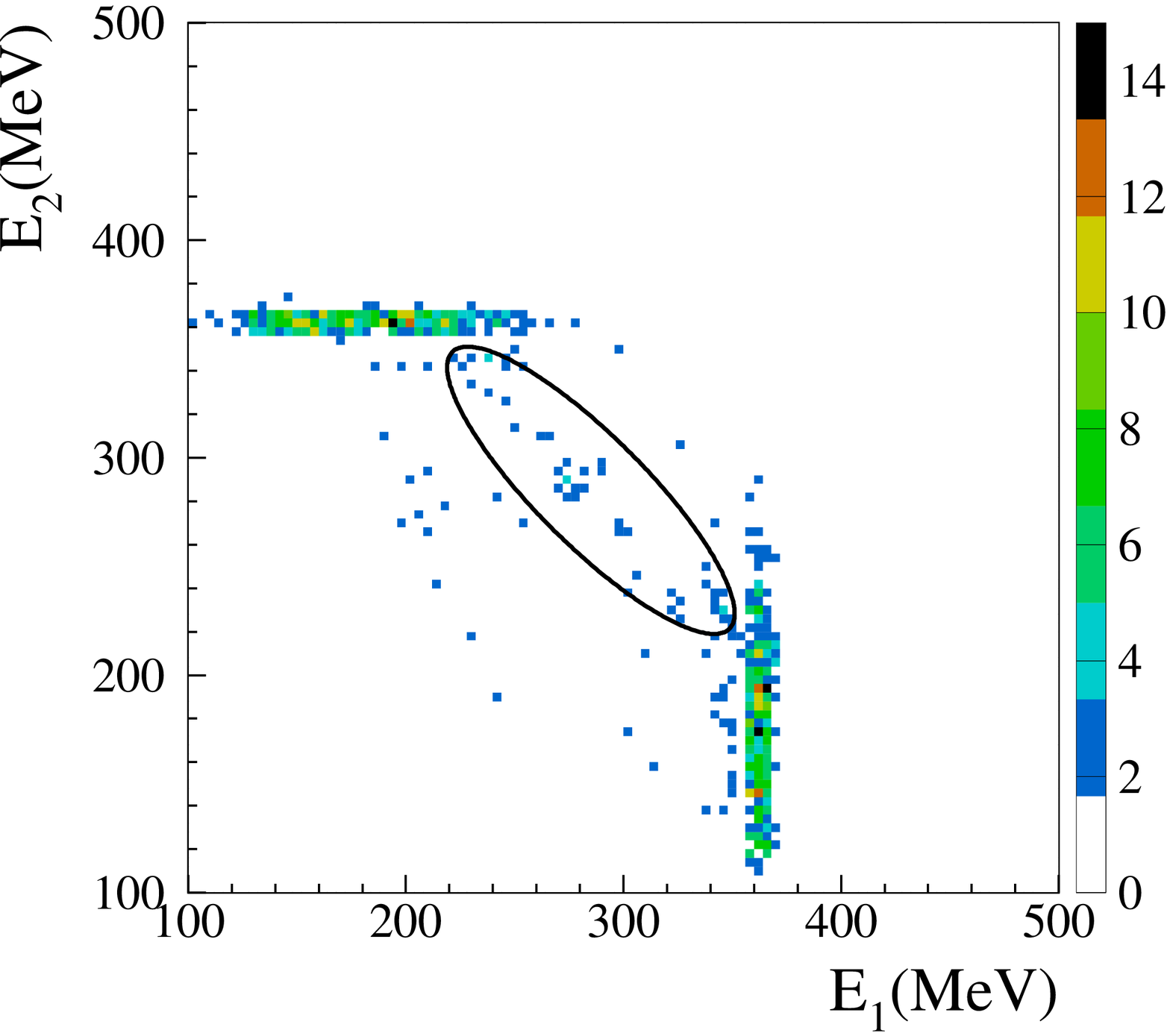,width=7cm}}
\raisebox{72ex}[0cm]{ \hspace{130pt} a) \hspace{190pt} b)}
\raisebox{36ex}[0cm]{\hspace{235pt} c)}
\caption{{\footnotesize Event density distributions in the $E_1-E_2$ plane. a) MC for $\eta'\gamma$ events; b) MC for $\eta\gamma$ events c)
Experimental data after first level selection; events in the $\eta\gamma$
bands have been downscaled for clarity. The number of observed events
inside the ellipse is 175.}}
\label{etapsel}
\end{figure}

\begin{figure}[ht]
\begin{center}
\epsfig{file=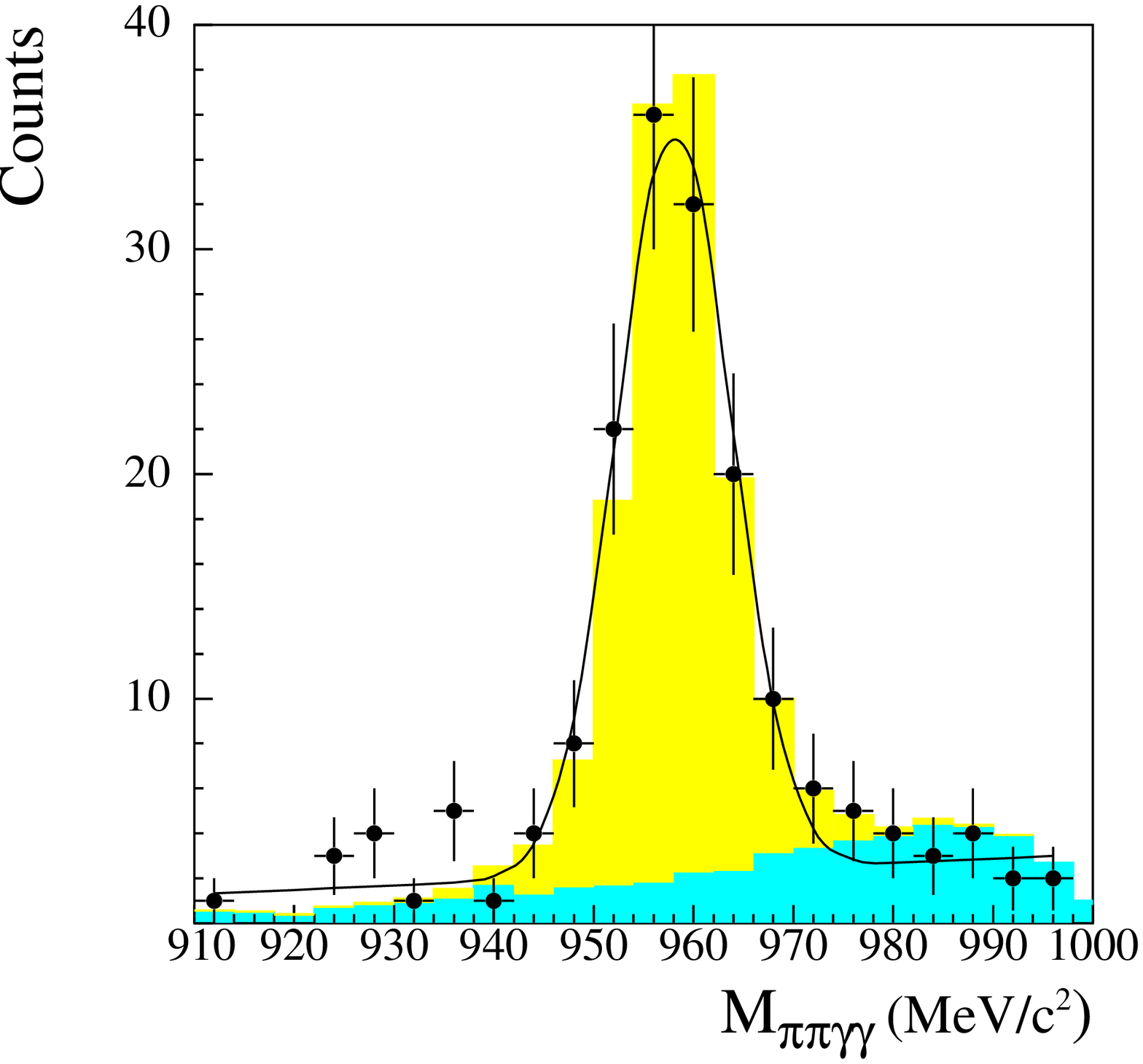,width=8cm}
\end{center}
\caption{{\footnotesize The $\pip\pim\phot\phot$ invariant mass for events selected as
$\fietapg$ candidates. The shaded areas represent signal (shape from MC) and
background (shape from sidebands analysis of data).The
continuous line is the result of a Gaussian plus linear fit.}}
\label{etap_peak}
\end{figure}

Process (1) events require, in addition to a)-c):
\bi
\item[f)] $E_{\pip}+E_{\pim}< 430 \MeV$ 
\item[g)]$\Sigma_{\gamma}E_{\gamma} > 540 \MeV$. 
\ei
These cuts are quite effective in suppressing events from process (3)
and (4) respectively. We estimate a residual background of less than 18 events from reactions 3 and 4. 
Contamination from process (2) is however still high. About 35\% of $\fietag$ events are still in the \fietapg\ sample for a S/B ratio of \ab\pt5,-3,.
To separate $\eta'\gamma$ and $\eta\gamma$ events we use the correlations
between the energies $E_1$ and $E_2$ of the two most energetic photons in
the event. Event densities in the $E_1$-$E_2$ plane are shown in fig.~\ref{etapsel}; for $\eta'\gamma$
events they are strongly anticorrelated (see MC distribution, fig.~\ref{etapsel}a) while for
$\eta\gamma$ events they are concentrated in two narrow bands around $E_1$
or $E_2=363$ MeV which is the energy of the radiative photon (see MC events in
fig.~\ref{etapsel}b). We select $\eta'\gamma$ candidate events inside the
elliptical shaped region as shown in fig.~\ref{etapsel}c for the
experimental data.
The $\pip\pim\gamma\gamma$ invariant mass for the events inside
the selection ellipse is plotted in fig.~\ref{etap_peak}. We notice a clear peak at the \etap\ mass value with the width expected from MC,
over a small residual background. 
The \etap\gam\ signal is obtained by a fit in the region $942\leq
M_{\pip\pim\gamma\gamma}\leq974$ MeV. For the signal we use the MC
shape. The background shape is derived from sidebands selected in the
$E1-E2$ plane just outside the acceptance ellipse.
The final number of events from process (1), after background subtraction, is $N_{\etap\gamma}=120\pm12({\rm stat.})\pm5({\rm syst.})$. 
The overall efficiency for $\eta'\gamma$ events is $\epsilon_{\etap\gamma}= 22.8\%$

The ratio of the branching ratios
$R=BR(\fietapg)/BR(\fietag)$ is determined from:
$$R={N_{\etap\gam}\over N_{\eta\gam}}\:
\left({\eps_{\eta\gam} \over \eps_{\etap\gam}}\right)\:
\frac{BR(\etapippimpiz)BR(\piz\rightarrow\gam\gam)}
{BR(\etappippimeta)BR(\etagg)}\;K_\rho$$
$K_{\rho}=0.95$ is a correction factor to the observed cross sections due to
the interference between the amplitudes $A(\f\to\eta(\etap)\gamma)$ and
$A(\rho\to\eta(\etap)\gamma)$ at $\sqrt(s)=m_{\phi}$. The correction factor $K_{\rho}$ has been evaluated
\cite{nota}, in a way similar to~\cite{Ampiezze}, using the
Gounaris-Sakurai~\cite{GS} parameterization of the $\rho$ and accounting
for the quark model phases which imply positive interference for $\eta'\gamma$
and negative interference for $\eta\gamma$ final state. 
Using 
the values in table \ref{tabsyst} we get
$$ R=\pt\left(4.70\pm0.47\ (\rm stat.)\pm0.31\ (\rm syst.)\right),-3,$$
\begin{table}[h]
\bc
\begin{tabular}{|l|l|l|}
\hline
Quantity & Value & Systematic error \\\hline
$N_{\etap\gam}/N_{\eta\gam}$  &  \pt2.39,-3, &  4.2\% (background) \\\hline
\parbox{.05\textwidth}{$$\frac{\eps_{\eta\gam}}{\eps_{\etap\gam}}$$}&
1.60 & \kern-6pt\begin{tabular}{ll}Preselection&2.2\%\\  
                          Photon counting & 0.8\% \\
                          Vertex efficiency  & 0.9\% \\
                  ${\rm Prob}(\chi^2)$ & 2.3\%\\
                          Accidentals        & 0.5\%\\
\end{tabular}\kern-6pt\\\hline
\parbox{.41\textwidth}{$${BR(\etapippimpiz)BR(\piz\rightarrow\gamma\gamma)
\over BR(\etappippimeta)BR(\etagg)}$$}&1.30&3.8\%\\\hline
TOTAL & & 6.6\%\\\hline
\end{tabular}\vglue3mm
\caption{{\footnotesize Contributions to the systematic error on $R$. The
systematics evaluation on the ratio of analysis efficiencies is obtained
from the study of the $\eta\gamma$ sample and varying the selection cuts. 
The intermediate BR's and errors are taken from \cite{PDG}.}}
\label{tabsyst}
\ec
\end{table}
\kern-5pt Systematics on luminosity and $\phi$ cross section
cancel out in the ratio exactly. Other effects, such as trigger and
reconstruction efficiencies, and machine background accidentals also
partially cancel out in evaluating $R$. The systematic error is
thus dominated by the uncertainties on background subtraction and on the
intermediate branching fractions \cite{PDG}.

Using the current PDG~\cite{PDG} value for BR($\fietag$) we 
extract the most precise determination of BR($\fietapg$) to date:
$$BR(\fietapg)=\pt6.10\pm0.61({\rm stat.})\pm0.43({\rm syst.}),-5,$$

\begin{figure}
\bc
\epsfig{file=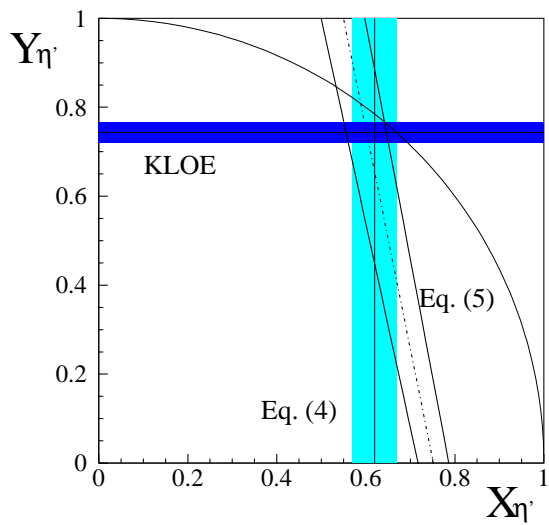,width=8cm}
\caption{{\footnotesize Bounds on $X_{\etap}$ and $Y_{\etap}$ from SU(3)
calculations and experimental branching fractions. 
The horizontal band is the KLOE result in the assumption $Z_{\etap}=0.$}}
\label{glue_cont}
\ec
\end{figure}

The value we obtain for $R$ can be related directly to the mixing angle in
the flavour basis. 
In the 
approach by Bramon {\it et al.} \cite{BraEsSca99} where SU(3) breaking is taken into account via
a constituent quark mass ratio $m_s/\overline m$ one has:
\begin{equation}
R =\cot^2\varphi_P\left(1-{m_s\over\overline m}\;\frac{\tan\varphi_V}{\sin
    2\varphi_P}\right)^2\!\left(\frac{p_{\eta'}}{p_{\eta}}\right)^3
\label{eqBramon}
\end{equation}
where $\fiv=3.4^{\circ}$ is the deviation from ideal mixing for
vector mesons and $p_{\eta(\etap)}$ is the radiative photon momentum in the
$\phi$ center of mass. 
Feldmann~\cite{Feld00} (following \cite{BallFreTyt96}) combines chiral anomaly predictions for $P\to\gamma\gamma$
with vector dominance to extract the couplings $g_{\phi\eta\gamma}$ and
$g_{\phi\etap\gamma}$. Then, apart from OZI rule violating terms:
\begin{equation}
R\!=\!\left({\sin\fip\sin\varphi_V\over 6f_q}-{\cos\fip\over 3f_s}\right)^2 
\kern-8pt\left/\kern-2pt
\left(\frac{\cos\fip\sin\varphi_V}{6f_q}+\frac{\sin\fip}{3f_s}\right)^2\!
\left(\frac{p_{\eta'}}{p_{\eta}}\right)^3\right.
\label{eqFeld}
\end{equation}
where $f_q$, and $f_s$ are the pseudoscalar decay constants in the flavour basis.
Values for all the parameters (except $\fip$) in equations \ref{eqBramon} and \ref{eqFeld} are from the quoted papers. Both approaches give very similar
results:
$$
\fip_{-(1)}=\left(41.8^{+1.9}_{-1.6}\right)^{\circ}\quad\hbox{and}\quad
\fip_{-(2)}=\left(42.2\pm1.7\right)^{\circ}$$
respectively. An estimate of the 
uncertainty on the extraction of $\fip$ using these approaches is
${\mathcal O}(0.5^{\circ})$. This is suggested by the difference among the
two values above and reflects the spread of the $m_s/\overline m$ , $f_s$  and $f_q$ values found in the literature.
The \fip\ value above is equivalent to a mixing angle of $\theta_P=
\left(-12.9^{+1.9}_{-1.6}\right)^{\circ}$ in the octet-singlet basis.
The mixing angle value has been obtained neglecting OZI rule violation and a possible gluonium contents of the $\eta$ and $\etap$ mesons.
Allowing for gluonium \cite{Ros83} we write:
\begin{equation}
\parbox{.8\textwidth}{$$\eqalign{
\ket{\eta}&=X_\eta\ket{u\bar u+d\bar d}/\sqrt{2}+Y_\eta\ket{s\bar s}+Z_\eta\ket{glue}\cr
\ket{\eta'}&=X_{\eta'}\ket{u\bar u+d\bar d}/\sqrt{2}+Y_{\eta'}\ket{s\bar s}+Z_{\eta'}\ket{glue}.\cr}$$}
\end{equation}
A gluonium component of the $\etap$ corresponds to $Z_{\etap}^2>0$ or equivalently
$X^2_{\etap}+Y^2_{\etap}<$1.
Constraints on $X_{\etap}$ and  $Y_{\etap}$ can be obtained in a
nearly model-independent way by using the following relations:
\begin{equation}
\label{X_etap}
\frac{\Gamma(\etap\rightarrow\rho\gamma)}{\Gamma(\omega\rightarrow\piz\gamma)}
\simeq
3\left(\frac{m^2_{\etap}-m^2_{\rho}}{m^2_{\omega}-m^2_{\pi}}\frac{m_{\omega}}{m_{\etap}}\right)^3
X_{\etap}^2
\end{equation}
and
\begin{equation}
\label{X_Y_etap}
\frac{\Gamma(\etap\rightarrow\gamma\gamma)}{\Gamma(\piz\rightarrow\gamma\gamma)}
=
\frac{1}{9}\left(\frac{m_{\etap}}{m_{\piz}}\right)^3(5X_{\etap}+\sqrt{2}Y_{\etap}\frac{f_{\pi}}{f_s})^2
\end{equation}
which are based on simple SU(3) ideas, exploiting the
magnetic dipole nature of the transitions V\to P\gam\ and P\to V\gam\ by deriving
the two photon couplings from the Wess-Zumino-Witten term of the
chiral Lagrangian~\cite{Ros83,BraEsSca01,Kou00}.
A consistency check of the assumption of $\eta -\eta'$ mixing without gluonium can
be performed as follows: if $Z_{\etap}=0$ one has $|Y_{\etap}|=\cos{\fip}$.  
This remains a reasonable approximation if the gluonium component is small.
In fig.~\ref{glue_cont} we plot in the $X_{\etap}$, $Y_{\etap}$ plane the
allowed bands corresponding to relations (\ref{X_etap}), (\ref{X_Y_etap}) and
to our measurement of $\cos{\fip}$ as well as the circumference
$X_{\etap}^2+Y_{\etap}^2=1$ corresponding to zero gluonium in the $\etap$. 
We thus find $Z_{\etap}^2=0.06^{+0.09}_{-0.06}$, compatible with zero
within 1$\sigma$ and consistent with a gluonium fraction below 15\%.

\section*{Acknowledgments}
We thank the DA$\Phi$NE team for their efforts in maintaining low
background running conditions and their collaboration during all
data-taking. We also thank Giuseppe~Fabio~Fortugno for his efforts in ensuring
good operations of the KLOE computing facilities. We thank R.~Escribano and
N.~Paver for fruitful discussions.
This work was
supported in part by DOE grant DE-FG-02-97ER41027; by 
EURODAPHNE, contract FMRX-CT98-0169; by the German Federal Ministry of Education and Research (BMBF) contract 06-KA-957; 
by Graduiertenkolleg 'H.E. Phys.and Part. Astrophys.' of 
Deutsche Forschungsgemeinschaft, Contract No. GK 742;
by INTAS, contracts 96-624, 99-37; and by TARI, contract HPRI-CT-1999-00088.


\begin{thebibliography}{99}
\bibitem{BeMor65} C.~Becchi and G.~Morpurgo  Phys. Rev {\bf 140},
B687 (1965)
\bibitem{BraGrauPan} A.~Bramon, A.~Grau and G.~Pancheri in ``The Second
DA$\Phi$NE Physics handbook'', vol II, 477 (ed.~L.~Maiani,
G.~Pancheri and N.~Paver, Frascati 1995) ;
E.~Marco, S.~Hirenzaki, E.~Oset and H.~Toki, Phys.\ Lett.\ {\bf B470}, 20
(1999)
\bibitem{Close92} F.~E.~Close ``Pseudoscalar mesons at DA$\Phi$NE'' in ``The
DA$\Phi$NE phyisics handbook'' vol.~II (ed.~L.~Maiani, G.~Pancheri and
N.~Paver, Frascati 1992)
\bibitem{Ben99} M.~Benayoun {\em et al.} Phys.\ Rev.\ {\bf D59}, 114027 (1999)
\bibitem{Ros83} J.~L.~Rosner  Phys. Rev. {\bf D27}, 1101 (1983)
\bibitem{BallFreTyt96} P.~Ball, J.-M.~Fr\`ere and M.~Tytgat
Phys.\ Lett.\  {\bf B365}, 367 (1996)
\bibitem{BraEsSca99}  A.~Bramon, R.~Escribano and M.~D.~Scadron  Eur. Phys
  J. {\bf C7}, 271 (1999)
\bibitem{BraEsSca01}  A.~Bramon, R.~Escribano and M.~D.~Scadron 
Phys.~Lett. {\bf B503}, 271 (2001)
\bibitem{Feld00} T.~Feldmann,  Int. Jou. Mod. Phys. {\bf A15}, 159 (2000)
\bibitem{KaisLeut98} H.~Leutwyler, Nucl.\ Phys.\ Proc.\ Suppl.\ {\bf 64},
223 (1998) [hep-ph/9709408] ; R.~Kaiser and H.~Leutwyler, hep-ph/9806336
\bibitem{EsFre99}
R.~Escribano and J.~M.~Frere,
Phys.\ Lett.\  {\bf B459}, 288 (1999)
\bibitem{FeldKroll98} Th.~Feldmann, P.~Kroll, and B.~Stech, 
Phys. Rev. {\bf D58}, (1998)
\bibitem{Defazio00} F.~De Fazio and M.~R.~Pennington,  JHEP 0007:051
(2000) 
\bibitem{expBKetap} CLEO Collaboration, Phys.\ Rev.\ Lett. {\bf 85}, 520
(2000); BABAR Collaboration  Phys.\ Rev.\ Lett. {\bf 87},
221802 (2001); BELLE Collaboration  Phys.\ Lett. \ {\bf B517},
309 (2001)
\bibitem{theoBKetap}
H.~Y.~Cheng and B.~Tseng,
Phys.\ Lett.\ B {\bf 415}, 263 (1997);
I.~Halperin and A.~Zhitnitsky,
Phys.\ Rev.\ D {\bf 56}, 7247 (1997);
A.~S.~Dighe, M.~Gronau and J.~L.~Rosner,
Phys.\ Rev.\ Lett.\  {\bf 79}, 4333 (1997);
A.~Ali, J.~Chay, C.~Greub and P.~Ko,
Phys.\ Lett.\ B {\bf 424}, 161 (1998);
M.~Ciuchini, R.~Contino, E.~Franco, G.~Martinelli and L.~Silvestrini,
Nucl.\ Phys.\ B {\bf 512}, 3 (1998)
M.~Z.~Yang and Y.~D.~Yang,
Nucl.\ Phys.\ B {\bf 609}, 469 (2001);
C.~Isola, M.~Ladisa, G.~Nardulli, T.~N.~Pham and P.~Santorelli
Phys.\ Rev.\ {\bf D65}, 094005 (2002)
\bibitem{Kou00} E.~Kou, Phys.\ Rev.\ D {\bf 63}, 054027 (2001)
\bibitem{PDG} The Particle Data Group (D.~Groom {\it et al.})
  Eur. Phys. Jou. {\bf C15} (2000)
\bibitem{SND99}
M.~N.~Achasov  {\it et al.} [SND Collaboration],
 JETP Lett. {\bf 69}, 97 (1999)
\bibitem{CMD200b}
R.~R.~Akhmetshin {\it et al.}  [CMD-2 Collaboration],
 Phys.~Lett. {\bf B473}, 337 (2000);
 Phys.~Lett. {\bf B494}, 26 (2000)
\bibitem{kloe} KLOE Collaboration, ``KLOE: a general purpose detector for
  DA$\Phi$NE'', LNF-92/019 (IR) (1992); KLOE Collaboration, ``The KLOE detector
  - Technical Proposal'', LNF-93/002 (IR) (1993).
\bibitem{dafne} S. Guiducci, Status of \dafne, in: P.~Lucas, S.~Webber,
Proc.\ of the 2001 Particle Accelerator Conference (Chicago, Illinois,
U.S.A.), 2001, pp. 353-355
\bibitem{DriftChamber} KLOE Collaboration, M.~Adinolfi {\it et al.},  ``The tracking detector of
the KLOE experiment'', 
to be published by Nucl.~Instr. and Meth. 
\bibitem{Calorimeter}
KLOE Collaboration, M.~Adinolfi {\it et al.}, 
 Nucl.~Instrum.~Meth. {\bf A482}, 364 (2002)
\bibitem{Trigger} KLOE Collaboration, M.~Adinolfi {\it et al.},  ``The KLOE
trigger system''
to be published by Nucl.~Instr. and Meth.
\bibitem{nota} F.~Ambrosino, KLOE note 179 ``Analysis of \fietapg\,
\fietag\ in \pip\pim\phot\phot\phot\ final state'' (May 2002)
\bibitem{Ampiezze}
R.~R.~Akhmetshin {\it et al.},
Phys.\ Lett.\ B {\bf 434}, 426 (1998).
\bibitem{GS} G.~J.~ Gounaris and J.~J.~Sakurai, Phys.\ Rev.\ Lett. {\bf
21}, 244 (1968)
\end{thebibliography}
\end{document}